\documentclass[preprint]{aastex}
\usepackage{graphics}
\usepackage{graphicx}
\usepackage[usenames]{color}

\newcommand{\ly}{Lyman-$\alpha$ }

\begin{document}

\title{Radiation Pressure in \ly Forest Clouds}

\author{Michael Fisher}
\affil{Department of Astronomy\\The Ohio State University\\140 West 18$^{\rm th}$ Avenue, Columbus, OH  43210-1173}
\affil{Current address:\\Battelle\\505 King Avenue, Columbus, OH  43201-2693}
\email{fisherml@battelle.org}

\begin{abstract}
The effective recombination coefficient, $\alpha_{\rm eff}$, is refined for optically thin cases. Radiation pressure in \ly and HeII \ly is calculated in terms of the number density and the mean free path. Pressure equilibrium between \ly clouds and an adiabatically expanding inter-galactic medium is assumed, $n_i T_i = n_c T_c$. Numerical models of isothermal and adiabatic expanding \ly forest clouds are presented, including evolving UV sources (QSOs), with various turn-on times $z_{on} = $ 20, 10, and 5, and with $q_0 = \frac{1}{2}$ in a matter-dominated Friedmann-Robertson-Walker Universe. These models lead to the conclusion that the radiation pressure and QSO turn-on time are significant in determining the range of physical size, $D$, and neutral hydrogen column density, $N(HI)$, permitted for stable \ly forest clouds.

This manuscript was written in 1989 and never submitted for publication.
\end{abstract}

\section{Introduction}
A great deal of effort is being expended on the absorption features of QSOs, especially the IGM clouds that comprise the \ly forest. Although the method of confinement for these clouds is still questionable, current research provides a detailed list of physical characteristics, including: physical size, column density and temperature, see \citet{carswell87}, Sargent et. al (1980) and Chaffee et. al (1986). Typical values for the physical size of the clouds is $0.4~kpc < D < 400~kpc$, Sargent (1988) and for neutral hydrogen column densities is $N(HI) \approx 10^{15}~cm^{-2}$. In addition to these physical characteristics, the \ly clouds exhibit a distribution in redshift and neutral hydrogen column density of the form
\begin{equation}
P(N_{H^0},z) \sim N_{H^0}^{-\beta} (1+z)^{\gamma}~~~,
\end{equation}
where $\beta \sim 1.7 \pm 0.2$ and $\gamma \sim 2.3 \pm 0.4$, and the expression is valid over the range of column densities $10^{13}~cm^{-2} \leq N_{H^0} \leq 10^{16}~cm^{-2}$, Carswell (1988) and Sargent (1988).

Numerical models of \ly forest clouds showing that the upper limit in column density for the distribution is due to radiation pressure are presented here. We adopt the quasar evolution model proposed by Schmidt \& Green (1983) with turn-on at $z =$ 20, 10, and 5, and follow the size and density evolution of the clouds that were stable to radiation pressure at the time the quasars formed. We show three models, A, B and C, by varying the magnitude of the background UV flux within accepted uncertainty. We also calculate an effective recombination coefficient, $\alpha_{\rm eff}$, for determining the fractional ionization of the cloud in the case $\tau << 1$ as a prelude to the model simulations.

\section{The Effective Recombination Coefficient}

\begin{figure}[t]
\begin{center}
\includegraphics{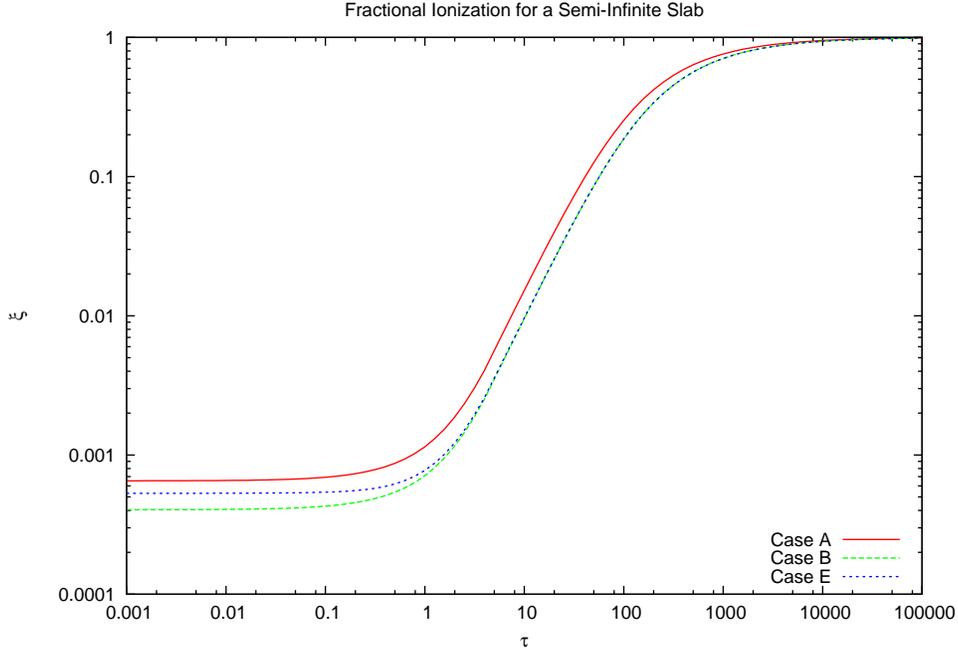}
\caption{The fractional ionization, $\xi = n_{HI} / n_{H}$, as a function of optical depth, $\tau_0$, for a semi-infinite medium with $T_e = 10^4~{\rm K}$. Case A uses the total recombination coefficient, Case B ignores recombinations to the ground state and Case E is the effective recombination coefficient.}
\label{fig1}
\end{center}
\end{figure}

The radiation field is normally separated into two parts, a ``source'' part, resulting from the background radiation field, and a ``diffuse'' part, resulting from the emission of the ionized gas, Osterbrock (1988). The ionization equation for a one element gas is then
\begin{equation}
N_{HI} \int_{\nu_0}^{\infty} \frac{J_{\nu}^{(s)} + J_{\nu}^{(d)}}{h \nu} \sigma_{\nu} d \nu = N_e N_p \alpha_A ~~~,
\end{equation}
where $\alpha_A$ is the total recombination coefficient. The ``source'' part is assumed to be a power law, $F_{\nu} \sim \nu^{-\alpha}$. For a plane parallel cloud of thickness $T_{\nu}$, the ``diffuse'' part is given by
\begin{equation}
\label{mihalas}
J_{\nu}^{(d)} =  \frac{8 \pi h}{c^2}\left( \frac{h^2}{2 \pi m_e k T_e} \right)^{\frac{2}{3}} e^{\beta}
\int_{0}^{T_{\nu}} \nu^3 e^{-\beta \frac{\nu}{\nu_0}} E_2(\tau_{\nu}) \frac{N_e N_p}{N_H} d \tau_{\nu} ~~~,
\end{equation}
where $\beta = h \nu_0 / k T_e$, Mihalas (1978). We then solve the ionization equation to obtain the fractional ionization with optical depth. For small optical depths, $\tau_0 < 1$, the fractional ionization lies between case A and case B, and approaches case B for large optical depths $\tau_0 \geq 1$, Figure~\ref{fig1}.

We desire a generalized numerical form of the efficiency factor that is valid for a variety of optical depths and electron temperatures. Therefore we solve the ionization equilibrium equation, including both source and diffuse terms, for different optical depths and temperatures. A simple function of optical depth, optical thickness and temperature is then fit to the solutions as the efficiency factor $\epsilon(\tau,T,T_e)$
\begin{equation}
\epsilon = \frac{\alpha_{{\rm eff}} - \alpha_B}{\alpha_1} ~~~.
\end{equation}

The general ionization equation is
\begin{equation}
\label{ioneq}
n_{HI} \left( \int_{\nu_0}^{\infty} \frac{J_{\nu}^{(s)}}{h \nu} \sigma_{\nu} d \nu + \int_{\nu_0}^{\infty} \frac{J_{\nu}^{(d)}}{h \nu} \sigma_{\nu} d \nu \right) = n_e n_p \alpha_A(T_e)~~~.
\end{equation}
The ionization state of the gas does not depend on the nature of the source of the ionizing radiation, only the number of ultraviolet photons greater than threshold. Therefore we take the source of ionizing radiation as an active galactic nucleus (AGN) characterized by a power law dependence on frequency throughout the ultraviolet portion of the continuum. This power law is characterized by spectral index $\alpha$ and the ionization parameter $\Gamma$
\begin{eqnarray}
F_{\nu} & \sim & \nu^{-\alpha} ~~~, \\
\Gamma & \sim & \frac{Q_0}{n_H c} ~~~,
\end{eqnarray}
where $Q_0$ is the number of ionizing quanta per unit area per unit time and $n_H$ is the total hydrogen density, $n_H = n_p + n_{HI}$.

The above form for the source term allows the first integral in Eq. (\ref{ioneq}) to be done analytically. Typical values for $\alpha$, $\Gamma$ and $Q_0$ are 3/2, $10^{-2}$ and $3 \times 10^{12}~{\rm cm^{-2}~s^{-1}}$, respectively. Then the number of photoionizations per unit volume per unit time due to the ``source'' radiation field is
\begin{eqnarray}
S & = & n_{HI} \int_{\nu_0}^{\infty} \frac{F_{\nu}^{(s)}(0)}{h \nu} \sigma_{\nu} e^{-\tau_{\nu}} d \nu ~~~, \\
& = & n_{HI} \sigma_0 \int_{1}^{\infty} Q x^{-11/2} e^{-\tau_0/x^3} dx ~~~, \nonumber
\end{eqnarray}
where $x=\nu/\nu_0$ and $Q_0 \equiv \int_{1}^{\infty} Q x^{-5/2} dx$. The substitution $y=\tau_0 x^{-3}$ allows the integral to be evaluated
\begin{equation}
S = Q_0 n_{HI} \sigma_0 \sum_{n=0}^{\infty} (-1)^n \frac{\tau_0^n}{n!} \frac{1}{2n+3} ~~~,
\end{equation}
which gives the number of photoionizations per unit volume per unit time at optical depth $\tau_0$ into the cloud. 

For $\tau_0 < 4$, the sum rapidly converges and in this form is easily used. For $\tau_0 \geq 4$, $\sum = 0.40157 \tau_0^{-1.4624}$ is an adequate approximation. The ionization with no ``diffuse'' part is 
\begin{equation}
n_{HI} Q_0 \sigma_0 \sum_{n=0}^{\infty} (-1)^n \frac{\tau_0^n}{n!} \frac{1}{2n+3} = n_e n_p \alpha_B ~~~,
\end{equation}
which is easily solved by letting $n_e = n_p$, and defining $\xi \equiv n_{HI}/n_H$. These substitutions result in a quadratic for the fractional ionization $\xi$
\begin{equation}
\xi^2 - \xi \left( 2+ \frac{c \Gamma \sigma_0}{\alpha_B} \sum_{n=0}^{\infty} (-1)^n \frac{\tau_0^n}{n!} \frac{1}{2n+3}  \right) + 1 = 0 ~~~.
\end{equation}

The ionization equation is now solved including the ``diffuse'' field by integrating Eq. (\ref{mihalas}) over frequency
\begin{equation}
n_{HI} \int_{\nu_0}^{\infty} \frac{J_{\nu}^{(d)}}{h \nu} \sigma_{\nu} d \nu = f(T_e) n_H^2 \int_{1}^{\infty} \frac{e^{-\beta x}}{x}dx \int_{0}^{\infty} E_2(\tau_{\nu}) [1-\xi(\tau_{\nu})]^2 d \tau_{\nu} ~~~,
\end{equation}
where $x=\nu/\nu_0$ and
\begin{equation}
f(T_e) = \frac{8 \pi \sigma_0 \nu_0^3 e^{\beta}}{c^2} \left( \frac{h^2}{2 \pi m_e k T_e} \right)^{2/3} ~~~.
\end{equation}
Using the definition for the exponential integral allows us to write the full ionization equation as 
\begin{eqnarray}
(1-\xi)^2 & = & \xi \frac{c \Gamma \sigma_0}{\alpha_A} \sum_{n=0}^{\infty} (-1)^n \frac{\tau_0^n}{n!} \frac{1}{2n+3} \\ \nonumber
&  & + \frac{f(T_e)E_1(\beta)}{\alpha_A} \int_0^{\infty} E_2(\tau_0)[1-\xi(\tau_0)]^2 d \tau_0 ~~~.
\end{eqnarray}

\begin{figure}[t]
\begin{center}
\includegraphics{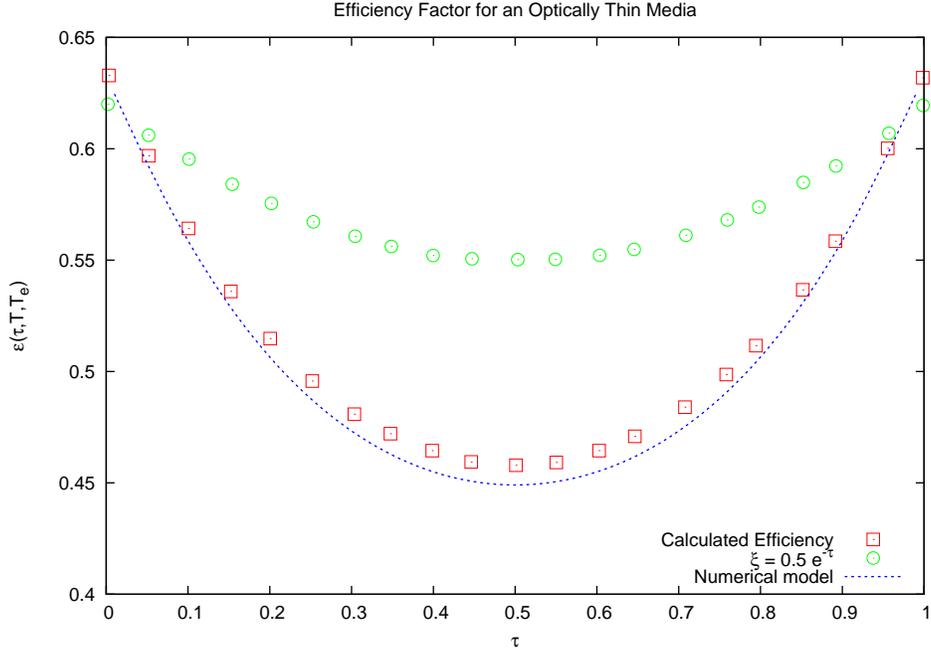}
\caption{The efficiency factor $\epsilon (\tau, T, T_e)$ vs optical depth for media of optical thickness $T=1$ and temperature $T_e = 10^4~{\rm K}$.}
\label{fig2}
\end{center}
\end{figure}

This equation is then solved for $\xi$ as a function of optical depth, $\tau_0$, shown as the calculated efficiency in Figure~\ref{fig2}.

These same results can be obtained by considering an ``effective'' recombination coefficient, $\alpha_{\rm eff} = \alpha_B + \epsilon(\tau,T,T_e) \times \alpha_1$, and considering only the ``source'' radiation field. The ``usual'' approximation for the efficiency has been $\epsilon = \case{1}{2} e^{-\tau}$. While this approximation works quite well for thick clouds, it does not accurately describe the situation for thin clouds as in the case of \ly forest clouds, Figure~\ref{fig2}. We present here a new form for the efficiency factor $\epsilon(\tau,T,T_e)$ dependent upon optical depth $\tau$, optical thickness $T$ and temperature $T_e$
\begin{equation}
\epsilon(\tau,T,T_e) = f(f+(1-g) \tanh y)((1- \tanh x) + (1- \tanh x'))~~~,
\end{equation}
where
\begin{eqnarray}
y & = & -0.3 \log T ~~~, \nonumber \\
x & = & 1.25(\log \tau + h) ~~~ \\
x' & = & 1.25(\log(T-\tau) + h) ~~~, \nonumber \\
t_e & = & T_e/10^3~~~. \nonumber
\end{eqnarray}
The expressions for $f$, $g$, and $h$ for are given in Table~\ref{fgh}. For $T_e > 10^5~ K$, the efficiency factor becomes
\begin{equation}
\epsilon = \inf[1,(1-0.007 \exp (-0.00338 t_e - 0.573 T))]~~~.
\end{equation}

\begin{table}
\begin{center}
\begin{tabular}{ccc}
\tableline \tableline
Factor & $T_e < 25,000 ~K$ & $25,000 < T_e < 10^5~K$ \\
\hline
$f$ & $0.2510 \exp(0.001518 t_e)$ & $(0.4346 + 0.0191 \log T_e)/2$  \\
$g$ & $0.9814 \exp(-0.002208 t_e)$ & $1.167+(0.0035-0.0144 \log T_e)\log T_e$ \\
$h$ & $0.3739 \exp(-0.006472 t_e)$ & $\log(3.3429 - 0.2794 \log T_e)$ \\ \tableline
\end{tabular}
\end{center}
\caption{Best fit for the factors $f$, $g$ and $h$.}
\label{fgh}
\end{table}

We used the photoionization code CLOUDY and our new model for the efficiency to reproduce the results from van Blerkom and Hummer (1967) as a check of the numerics.

\section{Evolution of the Background Ultraviolet Flux}

The number density of UV photons per unit frequency, $n(t,\nu)$, will satisfy the continuity equation
\begin{equation}
\frac{\partial n}{\partial t} + \nabla (n {\bf v}) + \frac{\partial}{\partial \nu} \left( n \frac{\partial \nu}{\partial t}\right) = S_\nu(t)~~~,
\end{equation}
where $v$ is the expansion velocity of the Universe and $S_{\nu}(t)$ is the source function for UV photons. Assuming that UV photons are continuously supplied by quasars, then $S_{\nu}(t)$ is proportional to the number of quasars. Following Ikeuchi \& Ostriker (1986), we adopt the quasar evolution model proposed by Schmidt \& Green (1983) with $q_0 = \frac{1}{2}$, $\beta = 9$, $z_{\rm on} = $ 20, 10, 5 and 2.5. Then the mean intensity at the Lyman limit is
\begin{equation}
4 \pi J_{\nu_T}(z) = 4 \pi h \nu_T c (1+z)^4 S_{\nu_T}(0) F(z,z_{\rm on})~~~,
\end{equation}
where we have assumed $S_{\nu}(0) \sim \nu^{-2}$ and $F(z,z_{\rm on})$ is defined by
\begin{equation}
F(z,z_{\rm on}) \equiv \int_z^{z_{\rm on}} \frac{\exp[\beta \tau(z)]}{H(1+z)^2}dz ~~~,
\end{equation}
with $\tau(z)$ as the fractional look-back time to the present age.

\begin{figure}[t]
\begin{center}
\includegraphics{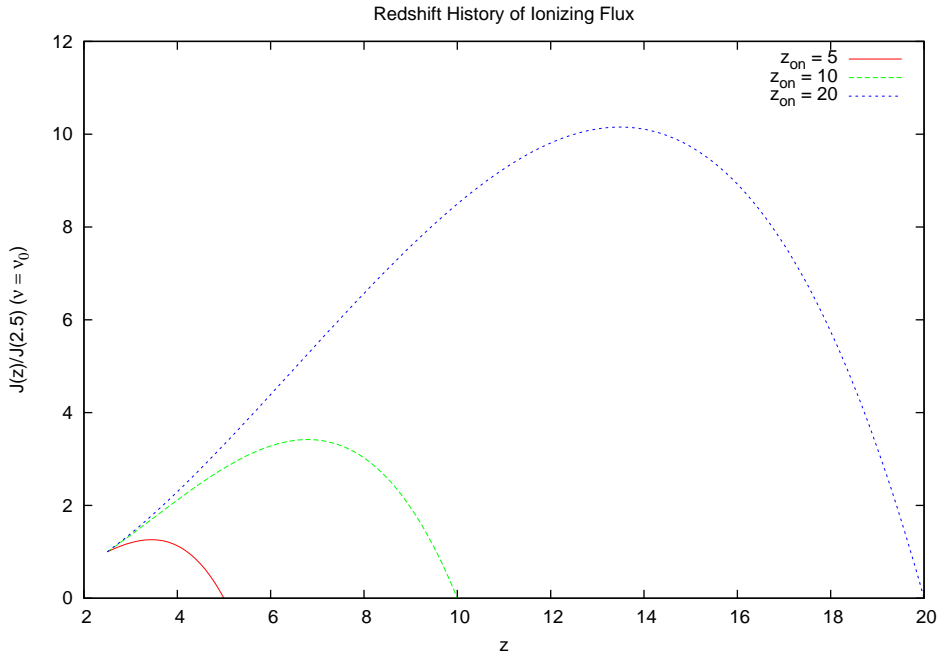}
\caption{The ratio of UV flux at redshift $z$ to UV flux at $z=2.5$ (at threshold) vs. redshift.}
\label{flux}
\end{center}
\end{figure}

\begin{figure}
\begin{center}
\includegraphics{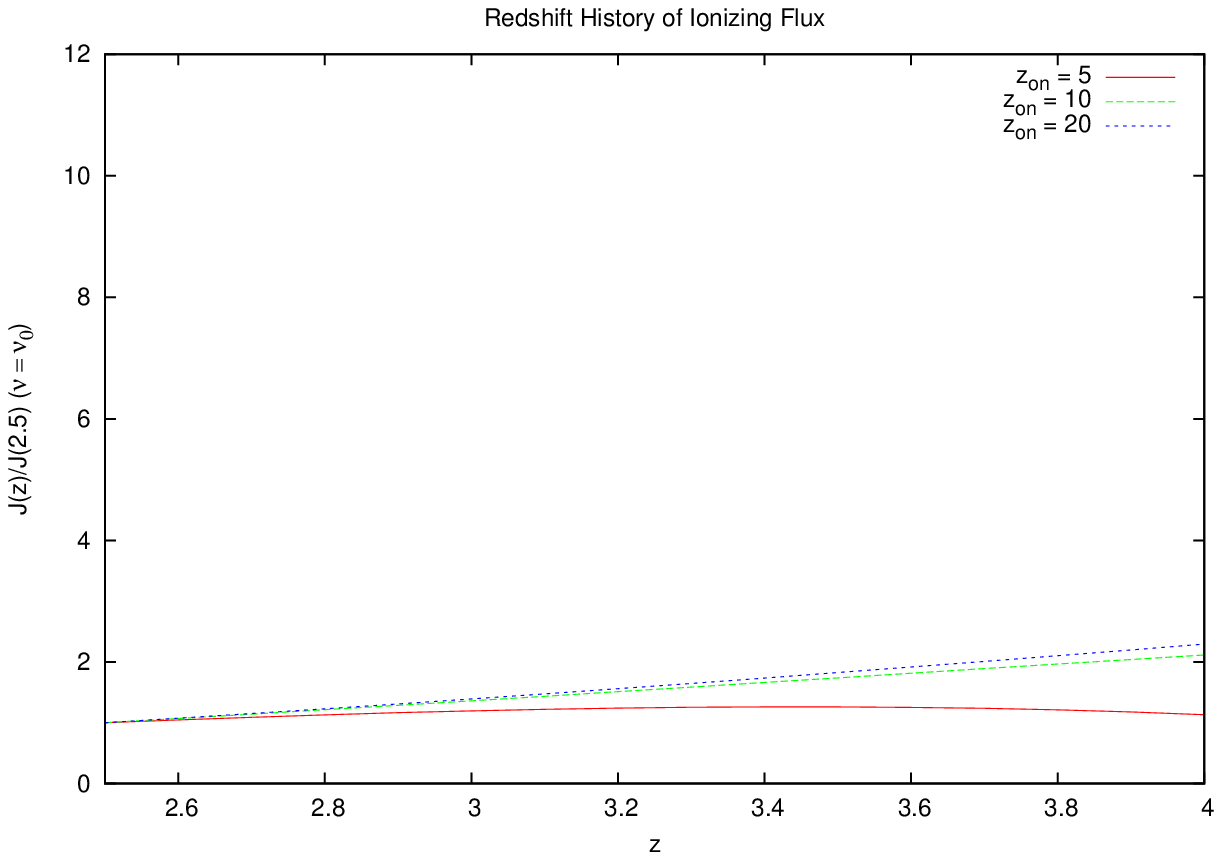}
\caption{The ratio of UV flux at redshift $z$ to UV flux at $z=2.5$ (at threshold) vs. redshift to redshift $z=4$, showing that $J_{\nu_T}$ is nearly constant for all three models  in this redshift range.}
\label{flux2}
\end{center}
\end{figure}

We define $S_{\nu_T}(0)$ such that $4 \pi J_{\nu_T}(2.5) = 4 \pi \times 10^{-21} ~ ergs ~ cm^{-2}~ s^{-1} ~ Hz^{-1} \times f$, Ostriker \& Ikeuchi (1983). We vary the factor $f$ in our models from 1 to 100 (models A, B and C). The ratio $J_{\nu_T}(z)/J_{\nu_T}(2.5)$ is shown in Figure~\ref{flux} for the evolutionary models described above. Figure~\ref{flux2} shows that the redshift range $2.5 \leq z \leq 4.0$, the Schmidt-Green QSO evolution model produces a background that is nearly constant and in agreement with Bajtlik, et. al (1988).

We let the IGM and the \ly clouds maintain pressure equilibrium and consider two cases for the expanding clouds: the clouds expand isothermally, or the clouds expand adiabatically. If the IGM is expanding adiabatically and the temperature of the IGM is non-relativistic, then
\begin{equation}
n_i(Z)T_i(z) = n_i(0)(1+z)^3 T_i(0) (1+z)^2 = n_i(0)T_i(0) (1+z)^5~~~.
\end{equation}
Then for isothermally expanding \ly clouds, $T_c(z)=T_c(0)$ and the density, $n_c$ and radius, $R_c$, for a fixed cloud mass scale as
\begin{equation}
n_c(z)=n_c(0)(1+z)^5 ~~~,
\end{equation}
and
\begin{equation}
R_c(z)= R_c(0)(1+z)^{-\frac{5}{3}} ~~~.
\end{equation}

For \ly clouds that expand adiabatically, the density scales as
\begin{equation}
n_c(z)=n_c(0)(1+z)^3~~~,
\end{equation}
and the radius of the cloud follows the expansion of the Universe
\begin{equation}
R_c(z)=R_c(0)(1+z)^{-1}~~~.
\end{equation}

\section{Radiation Pressure in the \ly Lines}

These primordial clouds are optically thin in the continuum, $\tau_0 \sim 10^{-2}$. However, they are optically thick in the \ly lines of hydrogen and helium, $\tau_{\alpha} \sim 10^2$, and therefore radiation pressure in these lines can be significant. The radiation pressure is $P_r = \frac{1}{3} h \nu_{\alpha} n_{\alpha}$, where $n_{\alpha}$ is the number density of the \ly photons. The number density is found by balancing production and destruction of photons, Mathews (1976) and Elitzur \& Ferland (1986). \ly photons are produced from recombinations to excited states
\begin{equation}
{\rm production} = n_e n_p \alpha_B~~~.
\end{equation}
These photons are destroyed by repeated scatterings until the photon escapes from the cloud. The number of photons that escape per unit volume per unit time, $n_{\rm esc}$, can be found from the number of \ly photons divided by the average time a photon spends within the cloud. This time scale is simply the mean free path divided by $c$. Therefore
\begin{equation}
{\rm destruction} = n_{\alpha} c / L ~~~,
\end{equation}
where $L$ is the mean free path. Bonilha et. al (1979) provide an analytical expression for the optical mean free path, $L_0$. To convert $L_0$ to mean free path, we must divide by the line absorption coefficient, $\alpha_0 = n_H \kappa_0$. Equating production and destruction and solving for $n_{\alpha}$
\begin{equation}
n_{\alpha} = \frac{L_0 n_e n_p \alpha_B}{c n_H \kappa_0}~~~,
\end{equation}
and for the radiation pressure
\begin{equation}
P_r = \frac{h L_0 n_e n_p \alpha_B}{3 \lambda_{\alpha} n_H \kappa_0}~~~.
\end{equation}

HeII \ly is treated differently since $\lambda$304 photons may also ionize H and therefore have an additional destruction mechanism. Bonilha et. al also give an expression for the optical mean free path in the presence of absorbers. Their $R$, which is the ratio of optical depth of absorbers, in this case $\tau_{912}$ and the optical depth at line center, $\tau_{304}$ is further reduced by the photoionization cross-section of hydrogen at 3 and 1 Rydbergs, $\sigma_3/\sigma_1$. Therefore their $\delta$ becomes
\begin{equation}
\delta = 3.704 \times 10^{-2} \frac{\tau_{912}}{\tau_{304}} L_0 ~~~,
\end{equation}
and the radiation pressure due to $\lambda$304 is
\begin{equation}
P_r = \frac{j L_0 n_e n_{HeIII} \alpha_{B}^{HeIII}}{3 \lambda_{304}n_{HeII} \kappa_{304} (1+0.9 \delta)^{0.97}} ~~~.
\end{equation}
In these models, radiation pressure is dominated by \ly and HeII \ly.

\section{Results}

The mean intensity, at threshold $J_{\nu}$, of the UV background is calculated at $z_{\rm on}$ and at $z=2.5$ for each model. The mean intensity is normalized at $z=2.5$ by: model A, $4 \pi J_{\nu}^A = 4 \pi \times 10^{-21}  ~ ergs ~ cm^{-2}~ s^{-1} ~ Hz^{-1}$; model B, $J_{\nu}^B = 10 \times J_{\nu}^A$; and model C, $J_{\nu}^C = 100 \times J_{\nu}^A$. For each model, and $z_{\rm on}$, the maximum cloud size before the cloud becomes unstable to radiation pressure ($P_r \geq P_g$) is calculated using the photoionization code CLOUDY. Finally a model is made of the cloud at $z=2.5$ scaling the density and radius of the maximum permissible cloud at $z_{\rm on}$ and the neutral column density is recorded. The density at $z_{\rm on}$ is varied and the procedure repeated. The results of our calculations are shown in Figures~\ref{fig4} and~\ref{fig5}. The area above and to the right of each curve is not allowed because of the instability due to radiation pressure at $z_{\rm on}$.

\begin{figure}[t]
\begin{center}
\includegraphics{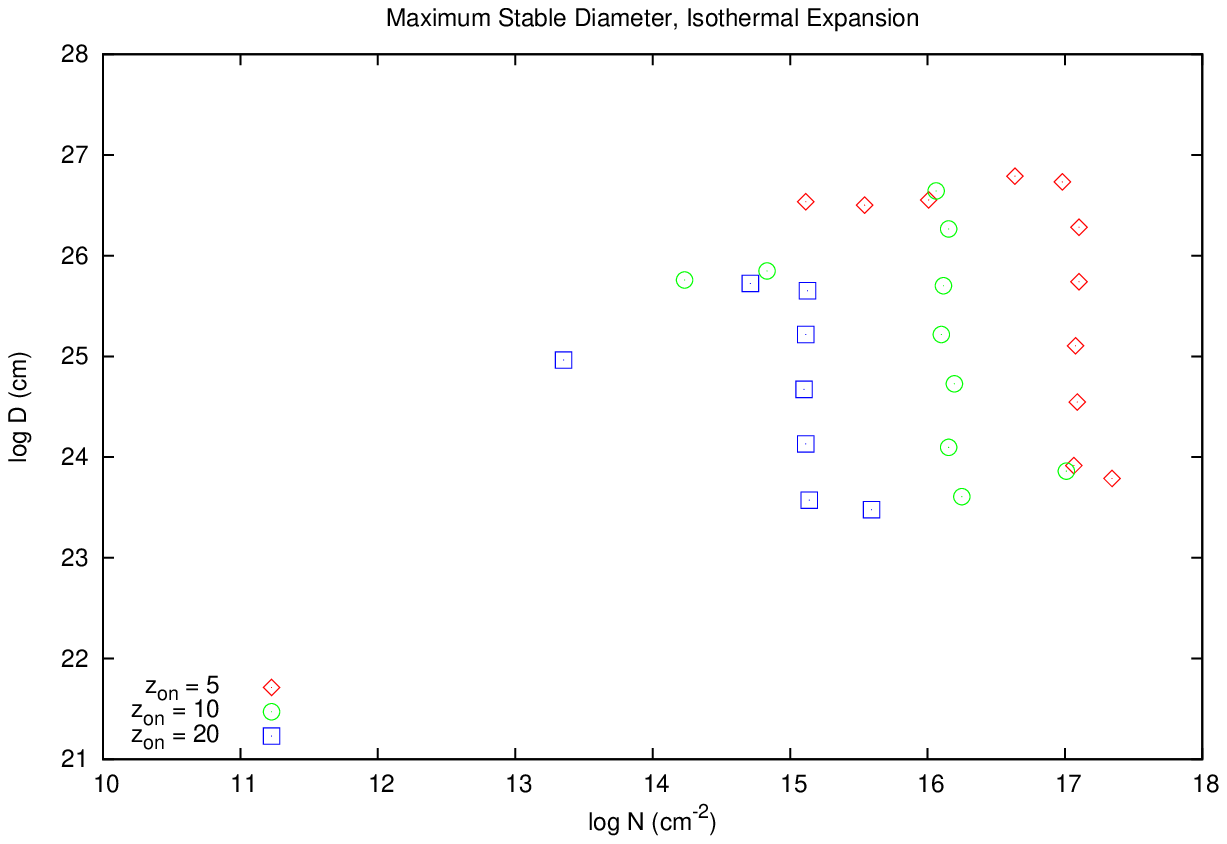}
\caption{Maximum stable diameter (in cm) vs neutral hydrogen column density (in cm$^{-2}$) for isothermally expanding clouds. $4 \pi J_{\nu_T}(2.5) = 4 \pi \times 10^{-21}~{\rm ergs~cm^{-2}~s^{-1}~Hz^{-1}}$.}
\label{fig5}
\end{center}
\end{figure}

\begin{figure}[t]
\begin{center}
\includegraphics{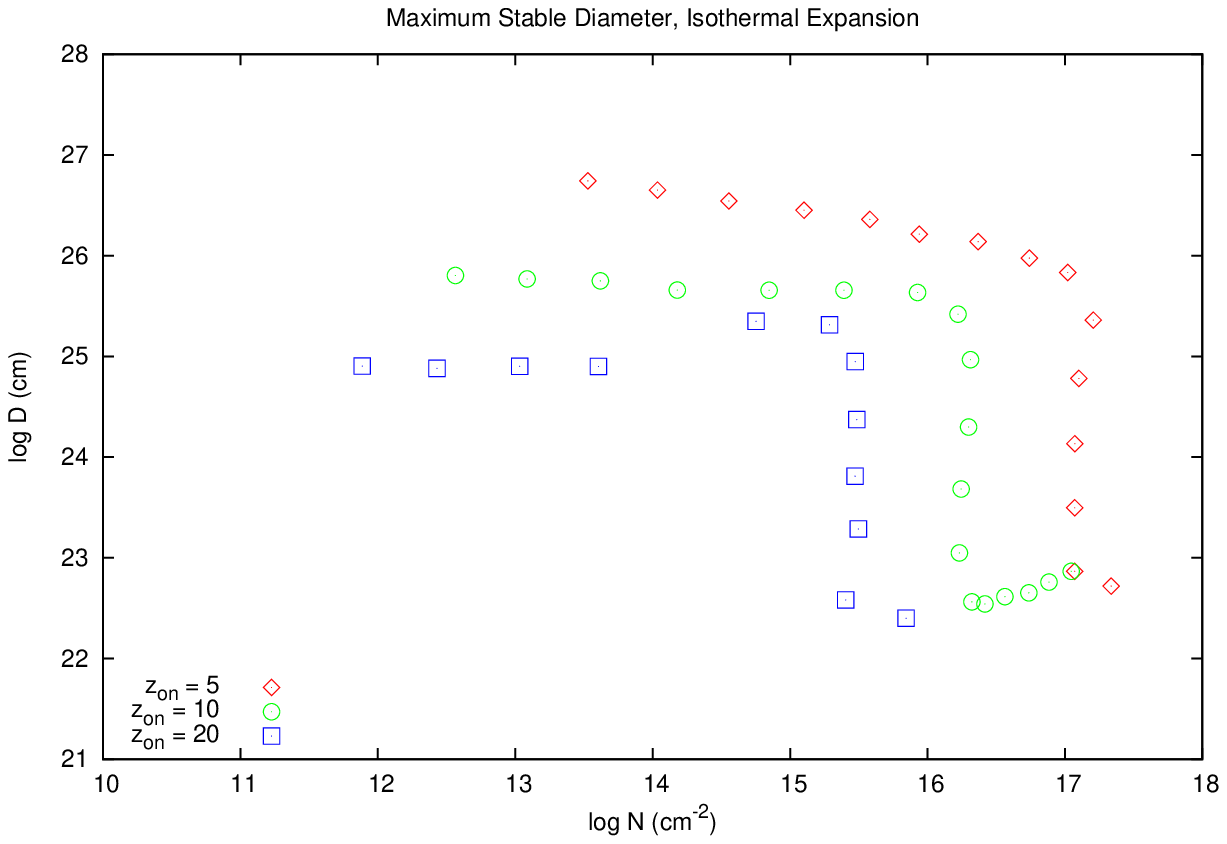}
\caption{Maximum stable diameter (in cm) vs neutral hydrogen column density (in cm$^{-2}$) for isothermally expanding clouds. $4 \pi J_{\nu_T}(2.5) = 4 \pi \times 10^{-20}~{\rm ergs~cm^{-2}~s^{-1}~Hz^{-1}}$.}
\label{fig5}
\end{center}
\end{figure}

There are two distinctive features in each model: a linear or nearly linear phase where the maximum permissible cloud size does not vary appreciably with increasing neutral hydrogen column density; and a vertical phase where the cloud size changes dramatically and the neutral hydrogen column density remains nearly constant despite changing the density of the cloud. It is noted that the appearance and location of the vertical phase is independent of the normalized value of the mean intensity, however the location in neutral hydrogen column density of the vertical phase {\em is} dependent upon the redshift chosen for $z_{\rm on}$ and the scaling law for cloud density. This is shown in Table~\ref{results}.
\begin{table}[t]
\begin{center}
\begin{tabular}{ccc}
\tableline \tableline
Expansion Method & $z_{\rm on}$ & $N_{HI}~(cm^{-2})$ \\ \tableline
& 5 & $10^{17}$ \\
Isothermal & 10 & $10^{16}$ \\
& 20 & $10^{15}$ \\ \tableline
& 5 & $10^{16}$ \\
Adiabatic & 10 & $10^{14}$ \\
& 20 & $10^{12}$ \\ \tableline
\end{tabular}
\end{center}
\caption{Results for isothermal and adiabatic expansion of \ly clouds.}
\label{results}
\end{table}

Metal free \ly systems are seen with neutral hydrogen column densities approaching $10^{16} ~ cm^{-2}$ and are {\em not} seen with $N_{HI} \geq 10^{16.5}~cm^{-2}$. We expect that the \ly clouds expansion is not strictly isothermal nor adiabatic, but lies in between these two extremes. Therefore the instability because of radiation pressure indicates that $z_{\rm on} \leq 10$ and the lack of metal free clouds with $N_{HI} \geq 10^{16.5} ~ cm^{-2}$ may indicate that $5 \geq z_{\rm on} \geq 10$.

With more observations of \ly forest clouds, definitive upper limits on the size of the clouds and on neutral hydrogen column density would decrease the uncertainty in the magnitude of the background UV flux and the epoch of $z_{\rm on}$ for QSOs.

\acknowledgments
The author gratefully acknowledges J.A. Baldwin, E.R. Capriotti, G.J. Ferland, J.V. Villumsen and R.J. Weymann for helpful dscussions and comments on the manuscript. This work was supported in part by National Science Foundation Grant AST 000-00.


\begin{thebibliography}{}

\bibitem[Atwood et al. (1985)]{atwood}
Atwood, B., Baldwin, J.A., and Carswell, R.F. 1985, \apj, 292, 58

\bibitem[Bajtlik et al. (1988)]{baj}
Bajtlik, S., Duncan, R.C., and Ostriker, J.P. 1988, \apj, 327, 570

\bibitem[Bechtold et al. (1987)]{bechtold}
Bechtold, J., Weymann, R.J., Lin, Z., and Malkan, M.A. 1987, \apj, 315, 180

\bibitem[Bergeron \& Stasi\'{n}ska (1986)]{berg}
Bergeron, J., and Stasi\'{n}ska, G. 1986, \aap, 169, 1

\bibitem[Bonilha et al. (1979)]{bon}
Bonilha, J.R.M., Ferch, R., Salpeter, E.E., Slater, G., and Noerdlinger, P.D. 1979, \apj, 233, 649

\bibitem[Carswell (1988)]{carswell88}
Carswell, R.F. 1988, in {\em QSO Absorption Lines: Probing the Universe} (eds. Blades, J.C., Turnshek, D., and Norman, C.A.) (Baltimore, MD: Space Telescope Science Institute), 91

\bibitem[Carswell et al. (1987)]{carswell87}
Carswell, R.F., Webb, J.K., Baldwin, J.A., and Atwood, B. 1987, \apj, 319, 709

\bibitem[Carswell et al. (1984)]{carswell84}
Carswell, R.F., Morton, D.C., Smith, M.G., Stockton, A.N., Turnshek, D.A., and Weymann, R.J., 1984, \apj, 278, 486

\bibitem[Chaffee et al. (1986)]{chaffee}
Chaffee, F.H., Jr., Foltz, C.B., Bechtold, J., and Weymann, R.J., 1986, \apj, 301, 116

\bibitem[Elitzur \& Ferland (1986)]{elitzur}
Elitzur, M., and Ferland, G.J., 1986, \apj, 305, 35

\bibitem[Ferland \& Rees (198?]{ferland}
Ferland, G.J., and Rees, M.J., 198?, \apj, 000, 0

\bibitem[Hunstead (1988)]{hunstead}
Hunstead, R.W., 1988, in {\em QSO Absorption Lines: Probing the Universe} (eds. Blades, J.C., Turnshek, D., and Norman, C.A.) (Space Telescope Science Institute), 71

\bibitem[Ikeuchi \& Ostriker (1986)]{Ikeuchi}
Ikeuchi, S., and Ostriker, J.P., 1986, \apj, 301, 522

\bibitem[Matthews (1976]{matthews}
Matthews, W.G., 1976, \apj, 207, 351

\bibitem[Mihalas (1978]{mihalas}
Mihalas, D., 1978, {\em Stellar Atmospheres}, 2nd Edition (W.H. Freeman: San Francisco)

\bibitem[Osterbrock (1988)]{osterbrock}
Osterbrock, D.E., 1988, {\em Astrophysics of Gaseous Nebulae and Active Galactic Nuclei}, (University Science Press)

\bibitem[Ostriker \& Ikeuchi (19830]{ostriker}
Ostriker, J.P., and Ikeuchi, S., 1983, \apjl, 268, L63

\bibitem[Sargent (1988)]{sargent}
Sargent, W.L.W., 1988, in {\em QSO Absorption Lines: Probing the Universe} (eds. Blades, J.C., Turnshek, D., and Norman, C.A.) (Space Telescope Science Institute), 1

\bibitem[Sargent et al. (1980)]{sargent2}
Sargent, W.L.W., Young, P.J., Boksenburg, A., and Tyler, D., 1980, \apjs, 42, 41

\bibitem[Steidel et al. (1988)]{steidel}
Steidel, C.C., Sargent, W.L.W., and Boksenberg, A., 1988, \apjl, 333, L5

\bibitem[Van Blerkom \& Hummer (1967)]{vanblerkom}
Van Blerkom, D., and Hummer, D.G., 1967, \mnras, 137, 353

\bibitem[Weymann et al. (1981)]{weymann}
Weymann, R.J., Carswell, R.J., and Smith, M.G., 1981, \araa, 19, 41

\end{thebibliography}
\end{document}